\documentclass[epj]{svjour}

\usepackage{graphics}

\begin{document}

\title{Counter-ions at single charged wall: Sum rules}

\author{Ladislav \v{S}amaj} 

\institute{Institute of Physics, Slovak Academy of Sciences,
D\'ubravska\'a cesta 9, 845 11 Bratislava, Slovakia}

\date{Received: date / Revised version: date}

\abstract{
For inhomogeneous classical Coulomb fluids in thermal equilibrium, 
like the jellium or the two-component Coulomb gas, there exists a variety of 
exact sum rules which relate the particle one-body and two-body densities. 
The necessary condition for these sum rules is that the Coulomb fluid
possesses good screening properties, i.e. the particle correlation functions 
or the averaged charge inhomogeneity, say close to a wall, 
exhibit a short-range (usually exponential) decay. 
In this work, we study equilibrium statistical mechanics of an electric double
layer with counter-ions only, i.e. a globally neutral system of equally
charged point-like particles in the vicinity of a plain hard wall 
carrying a fixed uniform surface charge density of opposite sign.
At large distances from the wall, the one-body and two-body counter-ion 
densities go to zero slowly according to the inverse-power law.
In spite of the absence of screening, all known sum rules are shown to hold 
for two exactly solvable cases of the present system: in the weak-coupling 
Poisson-Boltzmann limit (in any spatial dimension larger than one) and at 
a special free-fermion coupling constant in two dimensions.
This fact indicates an extended validity of the sum rules and provides 
a consistency check for reasonable theoretical approaches. 
} 

\maketitle

\section{Introduction} \label{sec:1}
One of relevant problems in soft condensed matter is the study of thermodynamic
properties of charged mesoscopic objects (colloids or macro-ions of several 
thousand elementary charges), immersed in a polar solvent such as water 
containing mobile micro-ions of low valence.
The charged object together with the surrounding micro-ions form a neutral 
entity, known as the electric double layer, which is of intense theoretical 
interest, see reviews \cite{Attard96,Hansen00,Boroudjerdi05,Messina09}.
Electric double layers are important also for predicting an effective
interaction between macro-ions in a solvent and, in particular, 
an ``anomalous'' like-charge attraction \cite{Guldbrand84,Kjellander84}. 

The general classical system of mobile micro-ions can be formulated 
in two ways.
The case ``counter-ions only'' corresponds to one species of equally
charged micro-ions which neutralize the macro-ion charge.
This is the situation when a mesoscopic object, dissolved in a polar solvent, 
acquires an electric charge through the dissociation of functional surface
groups, releasing in this way counter-ions into the solvent \cite{Hunter05}.
In the case ``added electrolyte'', there is an infinite reservoir of
$\pm$ charge pairs.

\begin{figure}[tbp]
\resizebox{0.38\textwidth}{!}{\includegraphics{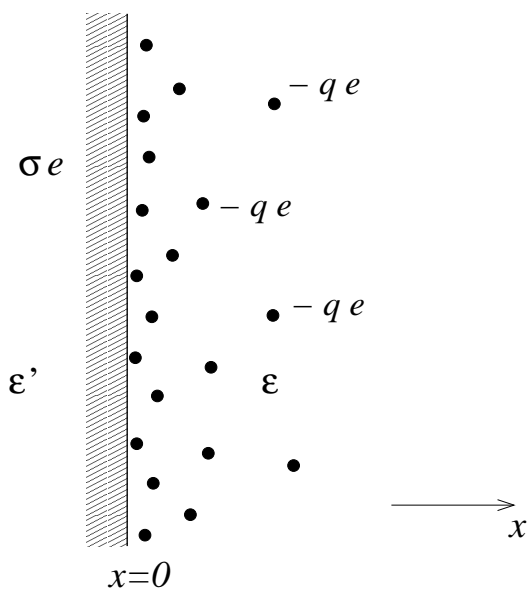}}
\caption{The general electric double layer with counter-ions only.
For simplicity, we consider the homogeneous dielectric case with 
$\epsilon'=\epsilon=1$ and the valence of counter-ions $q=1$.}
\label{fig:1}
\end{figure}

In this paper, we restrict ourselves to a single electric double
layer with counter-ions only, see the geometry in Fig. \ref{fig:1}. 
The system is $\nu$-dimensional ($\nu\ge 2$), inhomogeneous along 
the $x$-axis and translationally invariant along all remaining 
$\nu-1$ spatial axes.
The large macro-ion is approximated by a hard wall filled by a material
of dielectric constant $\varepsilon'$ in the half-space $x<0$.
Its surface at $x=0$ is charged by a fixed uniform surface-charge density 
$\sigma e$, $e$ being the elementary charge and say $\sigma>0$.
The wall is impenetrable to the $q$-valent counter-ions moving freely
in the complementary half-space $x>0$.
To simplify the notation, we set the valence $q=1$; to restore 
the $q$-dependence we have to substitute $\sigma\to \sigma/q$ and 
$e\to q e$ in final formulas. 
The counter-ions are classical point-like particles immersed in a solution 
of dielectric constant $\varepsilon$.
For simplicity, we consider the homogeneous dielectric case with 
$\varepsilon'=\varepsilon=1$ (vacuum in Gauss units), 
without electrostatic image forces.
The system is in thermal equilibrium at temperature $T$, or inverse
temperature $\beta=1/(k_{\rm B}T)$.
Although the dimension three (3D) is of primary physical interest, many
works deal with the two-dimensional (2D) version of Coulomb systems
interacting logarithmically which are sometimes exactly solvable,
at a specific free-fermion coupling 
\cite{Jancovici81,Jancovici84,Gaudin85,Cornu87,Cornu89}   
(for review, see \cite{Jancovici92}) or, in the case of the Coulomb gas of
point $\pm$ charges, in the whole stability interval of temperatures 
\cite{Samaj00,Samaj03}.

The model became, due to its relative simplicity, the cornerstone for
developing theoretical methods to general Coulomb systems. 
The weak-coupling (high-temperature) limit is described by the
Poisson-Boltzmann (PB) mean-field approach \cite{Andelman06} and by its 
systematic improvements via the loop expansion
\cite{Attard88a,Attard88b,Podgornik90,Netz00}.
The opposite strong-coupling (low-temperature) limit was pioneered by
Rouzina and Bloomfield \cite{Rouzina96} and developed subsequently in 
\cite{Grosberg02,Levin02}; a basic ingredient was the formation of 
a 2D Wigner layer of counter-ions close to the charged wall.
In the field-theoretical approach of Netz and collaborators
\cite{Moreira01,Netz01,Moreira02,Naji05}, the leading strong-coupling
behavior stems from a single-particle theory and the next correction orders 
correspond to a virial fluid expansion in inverse powers of the coupling 
constant. 
A comparison with Monte Carlo simulations shows the adequacy of the method
to capture the leading strong-coupling behavior, but its failure for the
first correction.
Recently \cite{Samaj11a,Samaj11b,Samaj11c}, a strong-coupling theory starting 
from the 2D Wigner crystal was proposed which reproduces the leading 
single-particle picture of the virial method and at the same time gives 
the first correction which is in excellent agreement with Monte Carlo data.

For dense homogeneous and inhomogeneous Coulomb fluids in thermal 
equilibrium, like the jellium or the two-component Coulomb gas, 
there exists a variety of exact sum rules which relate the particle 
one-body and two-body densities, for an old review see \cite{Martin88}. 
The necessary conditions for these sum rules is that the Coulomb fluid
exhibits good screening properties, i.e. the particle correlation functions 
or the averaged charge inhomogeneity, say close to a wall, exhibit a
short-range (usually exponential) decay to zero.
The bulk charge-charge correlation functions satisfy, in any dimension, 
the zeroth-moment and second-moment Stillinger-Lovett conditions
\cite{Stillinger68a,Stillinger68b}. 
In 2D, also the compressibility \cite{Vieillefosse75,Baus78} and higher-moment 
\cite{Kalinay00,Jancovici00} sum rules are available explicitly. 
For semi-infinite domains constrained by planar wall surfaces, the density of 
particles at the wall is related to the bulk pressure via the contact theorem 
\cite{Henderson78,Henderson79,Henderson81,Carnie81a,Wennerstrom82}.
The Carnie and Chan generalization to inhomogeneous fluids of 
the second-moment Stillinger-Lovett bulk condition gives the dipole sum rule
\cite{Carnie81b,Carnie83}. 
A sum rule for the variation of the particle density with respect to 
the external electrical field was derived by Blum et al. \cite{Blum81}. 
The WLMB (Wertheim, Lovett, Mou, Buff) equations \cite{Lovett76,Wertheim76}, 
originally derived for neutral particles, were adapted to charged systems 
as well \cite{Jancovici04}.
The charge-charge correlation functions decay slowly as an inverse-power
law along the wall \cite{Usenko76,Jancovici82a} and a sum rule for 
the amplitude function holds \cite{Jancovici82b,Jancovici95,Samaj10}.
A relation between this algebraic tail and the dipole moment was found 
in \cite{Jancovici01}.

The aim of this work is to investigate whether the known sum rules apply
to our electric double layer with counter-ions only.
The problem is that the present model does not exhibit good screening
properties: the density profile and the amplitude function of the
long-range tail along the wall exhibit a long-range (inverse-power) tail
at large distances from the wall.
We check the sum rules in two exactly solvable cases: the PB theory in any 
dimension $\nu\ge 2$ and at the special free-fermion coupling $\beta e^2 = 2$
in 2D. 
In both cases, all sum rules hold which indicates their extended applicability.
The sum rules provide exact information about the model and represent strong 
constraints which have to be fulfilled within a reasonable theoretical 
description.

The paper is organized as follows.
The known sum rules are listed in sect. \ref{sec:2}. 
The PB treatment in sect. \ref{sec:3} implies the inhomogeneous density 
profile; the correlation functions are evaluated with this profile by using 
a method which is alternative to the one-loop level of the field theory 
\cite{Netz00}.
The complete solution of the 2D model at the free-fermion coupling 
$\beta e^2 = 2$ is the subject of sect. \ref{sec:4}.
Sect. \ref{sec:5} is the Conclusion.

\section{List of sum rules} \label{sec:2}
We consider the homogeneous dielectric case $\varepsilon'=\varepsilon=1$
of the electric double layer pictured in Fig. 1.
Particles are constrained to the semi-infinite $\nu$-dimensional Euclidean 
domain. 
Each point of this domain ${\bf r}=(x,{\bf R})$ has the component 
$x\in [0,\infty]$, along which the system is inhomogeneous, and the remaining 
$\nu-1$ perpendicular (unbounded) components 
${\bf R}\in [-\infty,\infty]^{\nu-1}$, along which the system is 
homogeneous and translationally invariant. 

In dimension $\nu$, the Coulomb potential $v$ at point ${\bf r}$, 
induced by a unit charge at the origin ${\bf 0}$, 
is defined as the solution of the Poisson equation
\begin{equation} \label{Poisson}
\Delta v({\bf r}) = - s_{\nu} \delta({\bf r}) ,
\end{equation}
where $s_{\nu}=2\pi^{\nu/2}/\Gamma(\nu/2)$ (in particular, $s_2=2\pi$ and 
$s_3=4\pi$) is the surface area of the $\nu$-dimensional unit sphere.
Explicitly, 
\begin{equation}
v({\bf r}) = \left\{ 
\begin{array}{ll}
-\ln(r/r_0) &\mbox{if $\nu=2$,} \cr & \cr 
r^{2-\nu}/(\nu-2) &\mbox{otherwise,}
\end{array} \right.
\end{equation}
where $r=\vert {\bf r}\vert$ and $r_0$ is a length scale.
Dimension one has special features and is excluded from the analysis.
The definition of the Coulomb potential (\ref{Poisson}) implies 
the characteristic small-$k$ singularity $\hat{v}(k)=1/k^2$ 
in the Fourier space.
This maintains many generic properties, like screening and the corresponding
sum rules, of ``real'' 3D Coulomb systems with $1/r$ interactions.
Dimensions $\nu=2,3$ are of special physical interest.
The 2D logarithmic potential corresponds to the interaction of infinite 
uniformly charged lines which are perpendicular to the given plane.
Since the point-like particles possess the same charge, no short-range 
regularization of the Coulomb potential is needed and the thermodynamics 
is well defined.

The Hamiltonian of particles $i=1,\ldots,N$ with charge $-e$ at positions
$\{ {\bf r}_i=(x_i,{\bf R}_i) \}$ and the fixed surface charge density 
$\sigma e$ at $x=0$ reads
\begin{equation}
H = e^2 \sum_{i<j} v(\vert {\bf r}_i-{\bf r}_j\vert) + e^2 \sigma 
\frac{s_{\nu}}{2} \sum_i x_i .
\end{equation}  
In the canonical ensemble with the requirement of the overall neutrality,  
the thermal average at the inverse temperature $\beta$ will be denoted by
$\langle \cdots \rangle$.
At one-particle level, we introduce the particle density
\begin{equation}
n({\bf r}) \equiv n(x) = \left\langle \textstyle{\sum_i} 
\delta({\bf r}-{\bf r}_i) \right\rangle .
\end{equation}
At two-particle level, we have the two-body densities
\begin{eqnarray}
n_2({\bf r},{\bf r}') & \equiv & n_2(x,x';\vert {\bf R}-{\bf R}'\vert) 
= n_2(x',x;\vert {\bf R}-{\bf R}'\vert) \nonumber \\ 
& = & \left\langle \textstyle{\sum_{i\ne j}} 
\delta({\bf r}-{\bf r}_i) \delta({\bf r}'-{\bf r}_j) \right\rangle .
\end{eqnarray}
The corresponding (truncated) Ursell functions
\begin{equation}
n_2^{(\rm T)}({\bf r},{\bf r}') = n_2({\bf r},{\bf r}') - n({\bf r}) n({\bf r}')
\end{equation}
go to zero at large distances $\vert {\bf r}-{\bf r}'\vert\to\infty$.
In the above formulas, we used invariance with respect to translations
along the wall surface and rotations around the $x$ axis.

Now we list all known Coulomb sum rules, adapted to our system, 
forgetting for a while that the necessary conditions like screening 
do not apply.

\subsection{Sum rules involving the particle density only}
The overall electroneutrality of the system is equivalent to the condition
\begin{equation} \label{electro}
\int_0^{\infty} {\rm d}x\, n(x) = \sigma .
\end{equation}
Thus the particle density $n(x)$ must go to zero faster than $1/x$ 
as $x\to\infty$.

The contact theorem for planar wall surfaces 
\cite{Henderson78,Henderson79,Henderson81,Carnie81a,Wennerstrom82}
relates the contact density of particles $n(0)$ to the bulk pressure
of the fluid $P$ as follows
\begin{equation}
\beta P = n(0) - \frac{1}{2} \beta e^2 s_{\nu} \sigma^2 .
\end{equation}
Since the density of particles vanishes in the bulk $x\to\infty$, 
we have $P=0$ and the exact constraint
\begin{equation} \label{contact}
n(0) = \frac{1}{2} \beta e^2 s_{\nu} \sigma^2 .
\end{equation}

\subsection{Electroneutrality and dipole sum rules}
Two sum rules for inhomogeneous Coulomb fluids have the origin directly
in the zeroth-moment and second-moment Stillinger-Lovett conditions for 
the bulk correlation functions \cite{Stillinger68a,Stillinger68b}. 

The electroneutrality condition with a particle fixed at some point
takes the form
\begin{equation} \label{neutr}
n(x) = - \int_0^{\infty} {\rm d}x' \int {\rm d}{\bf R}\, n_2^{(\rm T)}(x,x';R) .
\end{equation}

The dipole sum rule
\begin{equation} \label{dipole}
\int_0^{\infty} {\rm d}x \int_0^{\infty} {\rm d}x' \int {\rm d}{\bf R}\, 
(x'-x) n_2^{(\rm T)}(x,x';R) = - \frac{1}{s_{\nu}\beta e^2}
\end{equation}
follows directly from the Carnie and Chan generalization to inhomogeneous 
fluids of the second-moment Stillinger-Lovett bulk condition 
\cite{Carnie81b,Carnie83}.
Note that the integrals over $x$ and $x'$ in (\ref{dipole}) cannot be 
interchanged since the integrated function is not absolutely integrable;
if the interchanging of integrations would be possible, the result is zero 
since $(x'-x)$ changes its sign under the exchange $x\leftrightarrow x'$
transformation.  

\subsection{Sum rule of Blum et al.}
Blum et al. \cite{Blum81} derived a sum rule which relates the variation
of the particle density $n({\bf r})$ with respect to the external electrical 
field $E=s_{\nu}\sigma e$ to the dipole moment with respect to a particle 
fixed at ${\bf r}$.
For our system, the sum rule reads explicitly as
\begin{equation} \label{Blum}
\frac{\partial n(x)}{\partial \sigma} = - s_{\nu} \beta e^2
\int_0^{\infty} {\rm d}x' \int {\rm d}{\bf R}\, (x'-x) n_2^{(\rm T)}(x,x';R) .
\end{equation}
This relation is superior to the dipole sum rule (\ref{dipole}) which
results by integrating both sides of (\ref{Blum}) over $x\in [0,\infty]$
and then considering the electroneutrality condition (\ref{electro}).

\subsection{The WLMB equations for charged systems}
The WLMB equations \cite{Lovett76,Wertheim76} 
were originally derived for neutral particles.
Their generalization to charged systems \cite{Martin88,Jancovici04}
relates the gradient of the density to an integral of the corresponding
two-body Ursell function over the boundary:
\begin{equation} \label{WLMB}
\frac{\partial n(x)}{\partial x} = \int {\rm d}{\bf R}\, n_2^{(\rm T)}(0,x;R) .
\end{equation}
Note that integrating this formula over $x\in [0,\infty]$ and
taking $n(x\to\infty)=0$, we recover the electroneutrality condition
(\ref{neutr}) for the special case $x=0$. 

\subsection{Long-range decay along the wall}
Because of asymmetry of the screening cloud around a particle sitting
near the wall, the two-body Ursell functions decay slowly along the wall 
\cite{Usenko76,Jancovici82a}.
Using linear response in combination with a simple macroscopic argument based 
on the electrostatic method of images \cite{Jancovici82b,Jancovici95,Samaj10},
one expects an asymptotic inverse-power law behavior
\begin{equation} \label{asym}
n_2^{(\rm T)}(x,x';R) \simeq \frac{f(x,x')}{R^{\nu}} , \qquad R\to\infty .
\end{equation}
In the $(\nu-1)$-dimensional Fourier ${\bf k}$-space with respect to 
${\bf R}$, defined by
\begin{eqnarray} 
\hat{g}(x,x';k) & = & \int {\rm d}{\bf R} 
\exp(-{\rm i}{\bf k}\cdot{\bf R}) g(x,x';R) , \nonumber \\ 
g(x,x';R) & = & \int \frac{{\rm d}{\bf k}}{(2\pi)^{\nu-1}} 
\exp({\rm i}{\bf k}\cdot{\bf R}) \hat{g}(x,x';k) , \label{Fourier}
\end{eqnarray}
this behavior is governed by the kink at ${\bf k}={\bf 0}$ \cite{Gelfand64}
of the small wave number behavior
\begin{equation} \label{smallk}
\hat{n}_2^{(\rm T)}(x,x';k) = \hat{n}_2^{(\rm T)}(x,x';0) 
- \frac{s_{\nu}}{2} f(x,x') k + \cdots . 
\end{equation}
  
The function $f(x,x')$, which is symmetric in $x$ and $x'$, 
obeys the sum rule \cite{Jancovici82b,Jancovici95,Samaj10}
\begin{equation} \label{tail}
\int_0^{\infty} {\rm d}x \int_0^{\infty} {\rm d}x'\, f(x,x') =
- \frac{2}{\beta e^2 s_{\nu}^2} .
\end{equation}
It is interesting that for all exactly solvable cases the function
$f(x,x')$ takes the product form  
\begin{equation} \label{productform}
f(x,x') = - g(x) g(x') .
\end{equation}
The decoupling of coordinates $x$ and $x'$ is intuitively due to
the fact that the lateral distance between the points ${\bf r}$
and ${\bf r}'$ goes to infinity and so the $x$-coordinates of the
two points become mutually uncorrelated. 

A relation between the algebraic tail of the correlation function
along the wall and the dipole moment of that function was 
found in Ref. \cite{Jancovici01}:
\begin{equation} \label{taildipole}
\int_0^{\infty} {\rm d}x' \int {\rm d}{\bf R}\, (x'-x) n_2^{(\rm T)}(x,x';R) 
= \frac{s_{\nu}}{2} \int_0^{\infty} {\rm d}x'\, f(x,x')
\end{equation}
for any $x\ge 0$.
The integration of this relation over $x\in [0,\infty]$ leads to the equality 
which is consistent with the previous sum rules (\ref{dipole}) 
and (\ref{tail}). 

\section{Poisson-Boltzmann theory} \label{sec:3}
We adapt to arbitrary $\nu\ge 2$ dimensions the derivation of the 3D PB theory 
by Andelman \cite{Andelman06} which was formulated originally longtime ago 
by Gouy \cite{Gouy10} and Chapman \cite{Chapman13}.

The density of counter-ions $n(x)$ corresponds to the charge density
$\rho(x)=-e n(x)$.
The average electrostatic potential $\phi$ at distance $x$ from the wall
satisfies the Poisson equation
\begin{equation}
\frac{{\rm d}^2 \phi(x)}{{\rm d}x^2} = - s_{\nu} \rho(x) .
\end{equation}
In the mean-field approach, the particle energy at $x$ is approximated by
$-e \phi(x)$ in the corresponding Boltzmann factor: 
\begin{equation}
n(x) = n_0 \exp\left[\beta e \phi(x)\right] . 
\end{equation}
The boundary condition for the electric field reads
\begin{equation}
\frac{{\rm d}\phi(x)}{{\rm d}x} \Big\vert_{x=0} =
- s_{\nu} \sigma e < 0 .
\end{equation}
We introduce the Gouy-Chapmann length
\begin{equation} \label{Gouy}
b = \frac{2}{\beta e^2\sigma s_{\nu}}
\end{equation}
and the dimensionless electric potential $\psi(x)=\beta e\phi(x)$, 
$n(x) = n_0 \exp[\psi(x)]$, which satisfies the equations
\begin{equation}
\frac{{\rm d}^2 \psi(x)}{{\rm d}x^2} = \frac{2 n_0}{\sigma b}
\exp\left[ \psi(x)\right] , \qquad
\frac{{\rm d}\psi(x)}{{\rm d}x} \Big\vert_{x=0} = - \frac{2}{b} .
\end{equation}
Their solution is 
\begin{equation}
\psi(x) = c - 2 \ln(x+b) , \qquad n_0 {\rm e}^c = \sigma b .
\end{equation}
The corresponding particle density
\begin{equation} \label{densityprof}
n(x) = \frac{\sigma b}{(x+b)^2}
\end{equation}
evidently satisfies both the electroneutrality condition (\ref{electro})
and the contact theorem (\ref{contact}).
Note that the spatial dimension $\nu$ enters only through the definition of 
the Gouy-Chapmann length (\ref{Gouy}).

The correlation functions can be evaluated with the density profile 
(\ref{densityprof}) by using the one-loop level of the field-theoretical 
method of Netz and Orland \cite{Netz00}.
The correlation functions are introduced there as auxiliary quantities 
to deduce the first correction to the density profile. 
Their exact meaning is not specified. 
This is why we propose an alternative derivation. 
It follows from the general theory of Coulomb fluids, 
namely the Ornstein-Zernike equation
\begin{equation} \label{OZ}
h({\bf r},{\bf r}') = c({\bf r},{\bf r}') + \int {\rm d}{\bf r}''
c({\bf r},{\bf r}'') n({\bf r}'') h({\bf r}'',{\bf r}') ,   
\end{equation} 
which relates the truncated pair correlation
\begin{equation}
h({\bf r},{\bf r}') = 
\frac{n_2^{(\rm T)}({\bf r},{\bf r}')}{n({\bf r}) n({\bf r}')}
\end{equation}
to the so-called direct correlation function $c$.
For particles interacting via the Coulomb potential $v$, the leading term of 
$c$ reads \cite{Jancovici82a,Kalinay00,Jancovici00}
\begin{equation}
c({\bf r},{\bf r}') \simeq - \beta e^2 v({\bf r},{\bf r}') .
\end{equation}
Inserting this expression into (\ref{OZ}), applying to both sides 
the Laplacian with respect to ${\bf r}$ and using the Poisson equation 
(\ref{Poisson}), we obtain
\begin{equation}
\left[ \Delta_{\bf r} - \beta e^2 s_{\nu} n(x) \right] h({\bf r},{\bf r}')
= \beta e^2 s_{\nu} \delta({\bf r}-{\bf r}') , \quad x,x'>0 
\end{equation}
and
\begin{equation}
\Delta_{\bf r} h({\bf r},{\bf r}') = 0 , \qquad x<0, x'>0 . 
\end{equation}
In the Fourier space with respect to ${\bf R}$ (\ref{Fourier}), we get
\begin{eqnarray} 
\left[ \frac{{\rm d}^2}{{\rm d}x^2} - \frac{2}{(x+b)^2} - k^2 \right]
\hat{h}(x,x';k) = \beta e^2 s_{\nu} \delta(x-x') , \phantom{aaa} \nonumber \\
x,x'>0 \phantom{aaa} \label{1e}
\end{eqnarray} 
and
\begin{equation} \label{2e}
\left[ \frac{{\rm d}^2}{{\rm d}x^2} - k^2 \right]
\hat{h}(x,x';k) = 0 , \quad x<0, x'>0 . 
\end{equation} 
These equations are supplemented by the conditions that $\hat{h}(x,x';k)$ and
${\rm d}\hat{h}(x,x';k)/{\rm d}x$ are continuous at $x=0$ and by regularity
$\hat{h}(x,x';k)\to 0$ as $x\to\pm\infty$. 
The formal solution of Eq. (\ref{1e}) reads \cite{Courant53} 
\begin{eqnarray}
\frac{\hat{h}(x,x';k)}{\beta e^2 s_{\nu}} & = & \frac{1}{W}
\left[ \varphi^+(x) \varphi^-(x') \theta(x-x') \right. \nonumber \\
& & \left. + \varphi^+(x') \varphi^-(x) \theta(x'-x) \right] 
+ A \varphi^+(x) \varphi^+(x) , \nonumber \\ & &
\end{eqnarray}
where $A$ is a free parameter, $\theta$ the Heaviside step function
\begin{equation}
\theta(x) = \left\{
\begin{array}{ll}
1& \mbox{if $x>0$,} \cr
0& \mbox{if $x<0$,}
\end{array} \right.
\end{equation} 
the functions $\varphi^{\pm}$ are given by
\begin{equation} \label{dif}
\left[ \frac{{\rm d}^2}{{\rm d}x^2} - \frac{2}{(x+b)^2} - k^2 \right]
\varphi^{\pm}(x) = 0 , \quad \varphi^{\pm}(\pm\infty) = 0 
\end{equation}
and the Wronskian
\begin{equation}
W = \varphi^-(x) \frac{\rm d}{{\rm d}x} \varphi^+(x) -
\varphi^+(x) \frac{\rm d}{{\rm d}x} \varphi^-(x)
\end{equation}
does not depend on $x$, as one can prove directly by differentiating
$W$ with respect to $x$ and then using differential equations (\ref{dif}) 
for $\varphi^{\pm}$.
The solutions of Eqs. (\ref{dif}) read
\begin{equation}
\varphi^{\pm}(x) = \left[ 1 \pm \frac{1}{k(x+b)} \right]
\exp\left[ \mp k(x+b) \right]
\end{equation} 
and the Wronskian $W = -2k$.
The formal solution of Eq. (\ref{2e}) is 
\begin{equation}
\frac{\hat{h}(x,x';k)}{\beta e^2 s_{\nu}} = B \varphi^+(x') \exp(k x) ,
\end{equation}
where $B$ is a free parameter.
The continuity conditions at $x=0$ determine the parameters $A$ and $B$
as follows
\begin{eqnarray}
A & = & - \frac{1}{2k} {\rm e}^{2kb} \frac{1}{1+2(kb)+2(kb)^2} , \nonumber \\ 
B & = & - \frac{1}{k} {\rm e}^{kb} \frac{(kb)^2}{1+2(kb)+2(kb)^2} .
\end{eqnarray}
After some algebra, the small-$k$ expansion of the truncated pair 
correlation is found to be
\begin{eqnarray}
\frac{\hat{h}(x,x';k)}{\beta e^2 s_{\nu}} & = & 
- \frac{1}{3} \frac{x_<^3 + 3 b x_<^2 + 3 b^2 x_< + 3 b^3}{(x+b)(x'+b)}
\nonumber \\ & & + \frac{b^4}{(x+b)(x'+b)} k + \cdots ,
\end{eqnarray}
where $x_< = \min(x,x')$.

The partial Fourier transform of the two-body density
$\hat{n}_2^{(\rm T)}(x,x';k) = n(x) n(x') \hat{h}(x,x';k)$
has the small-$k$ expansion
\begin{eqnarray} \label{exp}
\hat{n}_2^{(\rm T)}(x,x';k) & = & - \frac{4}{3\beta e^2 s_{\nu}} 
\frac{x_<^3 + 3 b x_<^2 + 3 b^2 x_< + 3 b^3}{(x+b)^3(x'+b)^3}
\nonumber \\ & & + \frac{4}{\beta e^2 s_{\nu}} 
\frac{b^4}{(x+b)^3(x'+b)^3} k + \cdots .
\end{eqnarray}
The first term of this expansion determines the integral
\begin{eqnarray}
& & \int {\rm d}{\bf R}\, n_2^{(\rm T)}(x,x';R) \nonumber \\ 
& & \qquad = - \frac{4}{3\beta e^2 s_{\nu}} 
\frac{x_<^3 + 3 b x_<^2 + 3 b^2 x_< + 3 b^3}{(x+b)^3(x'+b)^3} .
\end{eqnarray}
The presence of the second term in (\ref{exp}) signalizes the
asymptotic behavior of type (\ref{asym}).
With regard to formula (\ref{smallk}), the function $f(x,x')$ has
the long-range form
\begin{equation} \label{PBf}
f(x,x') = - \frac{8}{\beta e^2 s_{\nu}^2} \frac{b^4}{(x+b)^3(x'+b)^3} .
\end{equation}
Notice that the product form (\ref{productform}) takes place with 
$g(x) = b^2 \sqrt{8/(\beta e^2 s_{\nu}^2)}/(x+b)^3$. 
Using the last two relations it is easy to show that the sum rules
(\ref{neutr}), (\ref{dipole}), (\ref{Blum}), (\ref{WLMB}), (\ref{tail})
and (\ref{taildipole}) hold, in spite of the long-range nature of
the density profile $n(x)$ and the function $f(x,x')$.

\section{2D model at the free-fermion coupling} \label{sec:4}
The 2D version of the present model is mappable onto free fermions at 
the special coupling $\Gamma\equiv \beta e^2 = 2$ \cite{Jancovici84,Samaj11c}.
Here, we consider the Grassmann formulation of the Coulomb system on 
the surface of a semi-infinite cylinder of circumference $W$, i.e.
the strip with $x\in [0,\infty]$ and periodic boundary conditions along 
$y\in [0,W]$ \cite{Samaj11c}.
The charge line density $\sigma e$ at $x=0$ is neutralized by $N=\sigma W$
particles of charge $-e$.
At the coupling $\Gamma=2$, the renormalized one-body potential acting 
on particles reads
\begin{equation}
w_{\rm ren}(x) = \frac{4\pi}{W^2} \exp\left( - \frac{2\pi}{W} x \right) .
\end{equation}
In terms of two sets of anticommuting Grassmann variables $\xi_j$ and $\psi_j$,
defined on a discrete chain of $N$ sites $j=0,1,\ldots,N-1$,
the partition function of the system is expressible as 
\begin{equation}
Z_N = \int {\cal D}\psi {\cal D}\xi\, {\rm e}^{S(\xi,\psi)} , \qquad
S(\xi,\psi) = \sum_{j=0}^{N-1} \xi_j w_j \psi_j  ,
\end{equation}
where the diagonal interaction matrix has the elements
\begin{equation}
w_j = W \int_0^{\infty} {\rm d}x\, w_{\rm ren}(x) 
\exp\left( - \frac{4\pi}{W} j x \right) = \frac{1}{j+\frac{1}{2}} .
\end{equation}
Due to the diagonalized form of the anticommuting action $S(\xi,\psi)$,
the partition function is available explicitly as the product of
diagonal matrix elements
\begin{equation}
Z_N = w_0 w_1 \cdots w_{N-1} .
\end{equation}

Within the Grassmann formalism, the particle density is given by
\begin{equation} \label{for1}
n(x) = w_{\rm ren}(x) \sum_{j=0}^{N-1} \langle \xi_j\psi_j \rangle
\exp\left( - \frac{4\pi}{W} j x \right) ,
\end{equation}
where the two-correlator
\begin{equation} \label{for2}
\langle \xi_j\psi_j \rangle \equiv \frac{1}{Z_N}
\int {\cal D}\psi {\cal D}\xi\, {\rm e}^{S(\xi,\psi)} \xi_j \psi_j 
= \frac{\partial}{\partial w_j} \ln Z_N = \frac{1}{w_j} . 
\end{equation}
In the thermodynamic limit $N\to\infty$, which is equivalent to $W\to\infty$ 
keeping the ratio $N/W=\sigma$ fixed, we pass from the semi-infinite cylinder 
to a 2D half-space.
The variable $t=(j+\frac{1}{2})/(N-\frac{1}{2})$ becomes continuous and 
Eqs. (\ref{for1}), (\ref{for2}) imply the density profile
\begin{eqnarray}
n(x) & = & 4\pi \sigma^2\int_0^1 {\rm d}t\, t \exp(-4\pi\sigma t x) \nonumber \\
& = & 4\pi \int_0^{\sigma} {\rm d}s\, s \exp(-4\pi s x) \nonumber \\
& = & \frac{1}{4\pi x^2} \left[ 1 - \left( 1 + 4\pi \sigma x \right) 
{\rm e}^{-4\pi\sigma x} \right] . \label{2ddensity}
\end{eqnarray}
Like in the PB approach, the density has a long-range tail:
$n(x)\simeq 1/(4\pi x^2)$ as $x\to\infty$.
It is easy to verify that that the electroneutrality condition (\ref{electro})
and the contact theorem (\ref{contact}), taken with $\beta e^2=2$ and 
$s_2=2\pi$, hold.

The two-body densities are given by
\begin{eqnarray} 
n_2({\bf r},{\bf r}') & = & w_{\rm ren}(x) w_{\rm ren}(x') 
\sum_{j,k,j',k'=0\atop (j+j'=k+k')}^{N-1} \langle \xi_j \psi_k \xi_{j'} \psi_{k'} 
\rangle \nonumber \\ & & \times 
\exp\left( - \frac{2\pi}{W} [j z + k \bar{z} +j' z' + k' \bar{z}'] \right) , 
\label{for21}
\end{eqnarray}
where $z=x+{\rm i}y$ and $\bar{z}=x-{\rm i}y$ are the complex coordinates
of the point ${\bf r}=(x,y)$.
Due to the diagonal form of the action $S(\xi,\psi)$, the four-correlator
$\langle \xi_j \psi_k \xi_{j'} \psi_{k'}\rangle$ can be calculated using 
the Wick theorem, with the result
\begin{equation}
\langle \xi_j \psi_k \xi_{j'} \psi_{k'}\rangle = \frac{1}{w_j w_{j'}}
\left( \delta_{jk} \delta_{j'k'} - \delta_{jk'} \delta_{j'k} \right) .
\end{equation}
The first product of Kronecker functions $\delta_{jk} \delta_{j'k'}$ leads in 
(\ref{for21}) to the term $n(x) n(x')$ which has to be subtracted from 
$n_2({\bf r},{\bf r}')$ to obtain the truncated Ursell function. 
The second product $\delta_{jk'} \delta_{j'k}$ produces the Ursell function 
of the form
\begin{equation} \label{Ursell}
n_2^{(\rm T)}(x,x';R) = - n\left( \frac{x+x'+{\rm i}R}{2} \right)
n\left( \frac{x+x'-{\rm i}R}{2} \right) ,
\end{equation}
valid universaly for the finite-$W$ cylinder as well as for $W\to\infty$ 
2D half-space.
The integral representation of this formula for 2D half-space
\begin{eqnarray}
n_2^{(\rm T)}(x,x';R) & = & - (4\pi)^2 
\int_0^{\sigma} {\rm d}t\, t {\rm e}^{-2\pi t (x+x')}
\int_0^{\sigma} {\rm d}s\, s \nonumber \\ & & \times
{\rm e}^{-2\pi s (x+x')} {\rm e}^{-2\pi{\rm i}R(t-s)}
\end{eqnarray}
is useful to derive the important integral
\begin{equation}
\int_{-\infty}^{\infty} {\rm d}R\, n_2^{(\rm T)}(x,x';R)
= - (4\pi)^2 \int_0^{\sigma} {\rm d}t\, t^2 {\rm e}^{-4\pi t (x+x')} .
\end{equation}
Using this formula, the sum rules (\ref{neutr}), (\ref{dipole}), 
(\ref{Blum}) and (\ref{WLMB}) follow immediately.
The explicit representation of the relation (\ref{Ursell}) for 2D half-space
\begin{eqnarray}
n_2^{(\rm T)}(x,x';R) & = & - \frac{1}{\pi^2[(x+x')^2+R^2]^2} \nonumber \\
& & \times \Big\{ 1 - 2\left[ (1+2\pi\sigma(x+x')) \cos(2\pi\sigma R) 
\right. \nonumber \\ & & \left.
+ 2\pi\sigma R \sin(2\pi\sigma R) \right] \exp[-2\pi\sigma(x+x')]
\nonumber \\ & & + \left[(1+2\pi\sigma(x+x'))^2 +(2\pi\sigma R)^2 \right]
\nonumber \\ & & \times \exp[-4\pi\sigma(x+x')] \Big\} 
\end{eqnarray}
is convenient to specify the asymptotic $R\to\infty$ tail.
We obtain the expected $\nu=2$ behavior (\ref{asym}) with the function
\begin{equation} \label{2df}
f(x,x') = - 4 \sigma^2 \exp[-4\pi\sigma(x+x')] 
\end{equation}
which is again of the product form (\ref{productform}) with 
$g(x)=2\sigma\exp(-4\pi\sigma x)$.
In contrast to the PB result (\ref{PBf}), this function is short-ranged
with the exponential decay governed by the surface charge density.
The sum rules (\ref{tail}) and (\ref{taildipole}) are fulfilled.

\section{Conclusion} \label{sec:5}
The model of our present interest was the electric double layer with 
counter-ions only.
Such system is ``sparse'' in the sense that the particle density
and two-body densities vanish at asymptotically large distances 
from the wall.
Moreover, the large-distance decay is usually not fast enough as required 
by the linear response argument in the derivation of standard sum rules 
for classical inhomogeneous Coulomb fluids.

We studied two exactly solvable cases of the model. 
In the Poisson-Boltzmann limit, the long-range behavior takes place 
regardless of the dimension $\nu\ge 2$, for both the density profile
(\ref{densityprof}) and the asymptotic amplitude function (\ref{PBf}).
As concerns the 2D version of the model at the free-fermion coupling
$\Gamma=2$, the density profile (\ref{2ddensity}) has the long-range tail 
while the asymptotic function (\ref{2df}) is short-ranged.

The fact that the known sum rules are reproduced in two exactly solvable
cases of the electric double layer indicates their extended applicability.
The dipole sum rule (\ref{dipole}) deserves a special attention:
the tendency of the system to its bulk regime with the second-moment 
Stillinger-Lovett condition was crucial in its derivation.
As was mentioned before, the bulk regime of our model is ``emptiness''.

The asymptotic amplitude functions (\ref{PBf}) and (\ref{2df}) decouple 
themselves in $x$ and $x'$ coordinates, in analogy with the jellium and 
Coulomb-gas models \cite{Jancovici04,Jancovici82b}. 

We omitted from discussion ``dense'' systems like the one-component plasma 
with a neutralizing background (jellium) or the two-component Coulomb gas
of $\pm$ charges.
These systems exhibit exponential screening and so the sum rules automatically 
hold; their verification in special cases was given e.g. in 
\cite{Jancovici92,Jancovici04,Jancovici82b}. 

The extended validity of general sum rules for inhomogeneous Coulomb fluids
can serve as a useful check for the adequacy of weak-coupling 
physical theories.  
The strong-coupling theories like \cite{Samaj11a,Samaj11b,Samaj11c}
are based on the Wigner ground-state lattice structure formed by counter-ions.
The Coulomb system is not in its fluid phase in that regime and 
the sum rules do not longer apply.

\begin{acknowledgement}
The support received from Grant VEGA No. 2/0049/12 is acknowledged.
\end{acknowledgement}

\end{document}